\documentclass[final,5p,times,twocolumn]{elsarticle}

\usepackage{hyperref}

\usepackage{blindtext}

\usepackage{eqexpl}
\usepackage{amsmath}
\usepackage{amsfonts}
\usepackage{amssymb}
\usepackage{graphicx}
\usepackage{caption}
\usepackage{subfig}
\usepackage{mathtools}
\eqexplSetIntro{where:} 
\eqexplSetDelim{=} 

\journal{Nuclear Instruments and Methods A}

\begin{document}

\begin{frontmatter}

\title{Timing Performance of a Micro-Channel-Plate Photomultiplier Tube}

\author[CERN]{J.~Bortfeldt\fnref{lmu}}
\author[CERN]{F.~Brunbauer}
\author[CERN]{C.~David}
\author[CEA]{D.~Desforge}
\author[NSCR]{G.~Fanourakis}
\author[LIPP]{M.~Gallinaro}
\author[HIP]{F.~Garc\'ia}
\author[CEA]{I.~Giomataris}
\author[LIDYL]{T.~Gustavsson}
\author[CEA]{C.~Guyot}
\author[CEA]{F.J.~Iguaz\fnref{Iguaz}}
\author[CEA]{M.~Kebbiri}
\author[AUTH]{K.~Kordas}
\author[CEA]{P.~Legou}
\author[USTC]{J.~Liu}
\author[CERN]{M.~Lupberger\fnref{Lupberger}}
\author[AUTH]{I.~Manthos}
\author[CERN]{H.~M\"uller}
\author[AUTH]{V.~Niaouris}
\author[CERN]{E.~Oliveri}
\author[CEA]{T.~Papaevangelou}
\author[AUTH]{K.~Paraschou}
\author[LIST]{M.~Pomorski}
\author[CERN]{F.~Resnati}
\author[CERN]{L.~Ropelewski}
\author[AUTH]{D.~Sampsonidis}
\author[CERN]{T.~Schneider}
\author[CEA]{P.~Schwemling}
\author[LIST]{E.~Scorsone}
\author[CEA]{L.~Sohl\corref{cor}}
\ead{lukas.sohl@cern.ch}
\author[CERN]{M.~van~Stenis}
\author[CERN]{P.~Thuiner}
\author[NTUA]{Y.~Tsipolitis}
\author[AUTH]{S.E.~Tzamarias}
\author[RD51]{R.~Veenhof\fnref{Veenhof}}
\author[USTC]{X.~Wang}
\author[CERN]{S.~White\fnref{Virginia}}
\author[USTC]{Z.~Zhang}
\author[USTC]{Y.~Zhou}

\cortext[cor]{Corresponding author}

\address[CERN]{European Organization for Nuclear Research (CERN), CH-1211 Geneve 23, Switzerland}
\address[CEA]{IRFU, CEA, Universit\'e Paris-Saclay, F-91191 Gif-sur-Yvette, France}
\address[NSCR]{Institute of Nuclear and Particle Physics, NCSR Demokritos, 15341 Agia Paraskevi, Attiki, Greece}
\address[LIPP]{Laborat\'orio de Instrumenta\c{c}\~ao e F\'isica Experimental de Part\'iculas, Lisbon, Portugal}
\address[HIP]{Helsinki Institute of Physics, University of Helsinki, 00014 Helsinki, Finland}
\address[LIDYL]{LIDYL, CEA-Saclay, CNRS, Universit\'e Paris-Saclay, F-91191 Gif-sur-Yvette, France}
\address[AUTH]{Department of Physics, Aristotle University of Thessaloniki, Thessaloniki, Greece}
\address[USTC]{State Key Laboratory of Particle Detection and Electronics, University of Science and Technology of China, Hefei 230026, China}
\address[LIST]{CEA-LIST, Diamond Sensors Laboratory, CEA-Saclay, F-91191 Gif-sur-Yvette, France}
\address[NTUA]{National Technical University of Athens, Athens, Greece}
\address[RD51]{RD51 collaboration, European Organization for Nuclear Research (CERN), CH-1211 Geneve 23, Switzerland}

\fntext[lmu]{Now at Ludwig-Maximilians-University Munich, Germany}
\fntext[Iguaz]{Now at Synchrotron Soleil, BP 48, Saint-Aubin, 91192 Gif-sur-Yvette, France}
\fntext[Lupberger]{Now at Physikalisches Institut, Universit\"at Bonn, Germany}
\fntext[Veenhof]{Also at National Research Nuclear University MEPhI, Kashirskoe Highway 31, Moscow, Russia;
and Department of Physics, Uluda\u{g} University, 16059 Bursa, Turkey.}
\fntext[Virginia]{Also at University of Virginia}

\begin{abstract}
The spatial dependence of the timing performance of the R3809U-50 Micro-Channel-Plate PMT (MCP-PMT) by Hamamatsu was studied in high energy muon beams. Particle position information is provided by a GEM tracker telescope, while timing is measured relative to a second MCP-PMT, identical in construction. 
In the inner part of the circular active area (radius r$<$5.5\,mm) the time resolution of the two MCP-PMTs combined is better than 10~ps.
The signal amplitude decreases in the outer region due to less light reaching the photocathode, resulting in a worse time resolution. The observed radial dependence is in quantitative agreement with a dedicated simulation. 
With this characterization, the suitability of MCP-PMTs as $\text{t}_\text{0}$ reference detectors has been validated.
\end{abstract}


\end{frontmatter}

\section{Introduction}
Reliable reference detectors with high time resolution are needed in the characterization of new detector technologies aiming at performing ultra precise time measurements. An example is the PICOSEC-Micromegas \cite{Bortfeldt:2017jb,Papaevangelou:2016knm} concept, a newly introduced Micropattern Gaseous Detector for fast timing applications. The time resolution of several detector prototypes has been studied in great detail using minimum ionizing particle (MIP) beams at the CERN SPS secondary beam lines. 

There are different types of reference timing detectors providing less than 10\,ps time resolution. One possible option are silicon based detectors like SiPMs that have shown a timing performance in this range \cite{BENAGLIA201630}. Another detector technology with good timing response are MCP-PMTs. Those detectors are commonly used in various fields. One example is the use for time-of-flight positron emission tomography (PET). Other studies have shown a coincidence time resolution (CTR) of 30 ps FWHM \cite{Ota2019}. In this work, two MCP-PMTs of type Hamamatsu R3809U-50 Micro-Channel-Plate Photomultiplier Tubes (MCP-PMT) \cite{mcpdata} have been studied for the beam test measurement of the PICOSEC-Micromegas. 

Particles enter the MCP-PMT, traverse a radiator and generate Cherenkov light, which is then converted to charge in a multialkali photocathode between the radiator and multichannel plate. The radiator consists of a 3.2\,mm thick synthetic silica window that is integrated in the MCP-PMT. The useable photocathode diameter is 11\,mm as indicated. No further information about the photocathode and the window is given by the manufacturer. For the further simulation a generic Cherenkov angle of 45$^\circ$ is assumed.

These MCP-PMTs provide short signals with a rise time\footnote{10\,\% to 90\,\% amplitude} of 160\,ps.
Signals of this kind are well suited as a time reference for fast-timing detector studies. As discussed in a recent review of state-of-the-art timing detectors \cite{Vavra:2017jv}, it is advantageous for beam measurements to use a reference detector with time resolution significantly better than that expected for the detector under study. 
Measurements demonstrated the suitability of MCP-PMTs as a ($\text{t}_\text{0}$) timing reference for PICOSEC-Micromegas fast-timing detectors~\cite{Sohl:2018ls}. In this manuscript, we discuss further investigations, aiming at understanding the spatial dependence of the time resolution over the surface of the photocathode. The propagation of Cherenkov light including reflection and absorption has been modeled analytically and with a Monte-Carlo simulation. The observed radial dependences of the mean signal charge and time resolution have been compared to the modeled fraction of Cherenkov light reaching the photocathode. In the following, the different models as well as the measurement set-up will be explained and the measured data will be compared to the results from the simulation.

\section{Modeling of Cherenkov Light Propagation in the Radiator}\label{sec:sim}

\begin{figure}[!htb]
	  	   		\centering
	  	   		\includegraphics[width =  \linewidth]{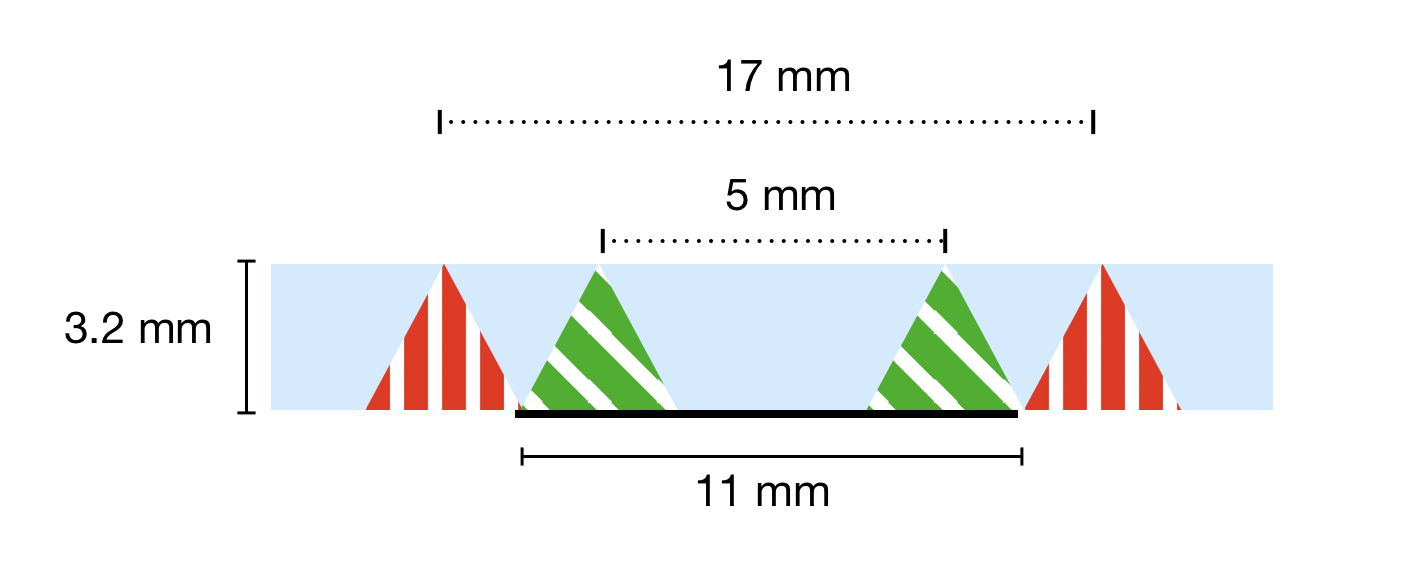}
	  	   		\caption[Cherenkov Window Sketch]{Schematic Cherenkov light cone propagation in the radiator, as assumed in the model. The green colored cones (diagonal hatching) are fully projected onto the photocathode, while the projection of the red coloured cones (vertical hatching) are fully outside of the photocathode area. The radiator has a thickness of 3.2 mm, and the photocathode a diameter of 11 mm.}
	  	   		\label{fig:window}
	  	   	\end{figure}

When a particle passes through the MCP-PMT at a larger distance from the photocathode center, less light reaches the photocathode as the Cherenkov cone is not fully projected onto the photocathode. Figure~\ref{fig:window} shows a sketch of the Cherenkov cones and the photocathode with the assumed dimensions of the detector. In our model, Cherenkov photons can be either converted to charge at the photocathode, or can be reflected or absorbed at the radiator boundaries. Part of the reflected light can later reach the photocathode and contribute to the detector signal. An analytic and a Monte-Carlo model have been developed to predict the amount of light reaching the photocathode. 

\subsection{Analytic Geometric Modeling}
A geometric calculation of the overlapping areas of photocathode and Cherenkov cone for particle impact points at different radii has been done, with and without considering reflection between the radiator and the photocathode.

\begin{figure}[!htb]
	  	   		\centering
	  	   		\includegraphics[width =  \linewidth]{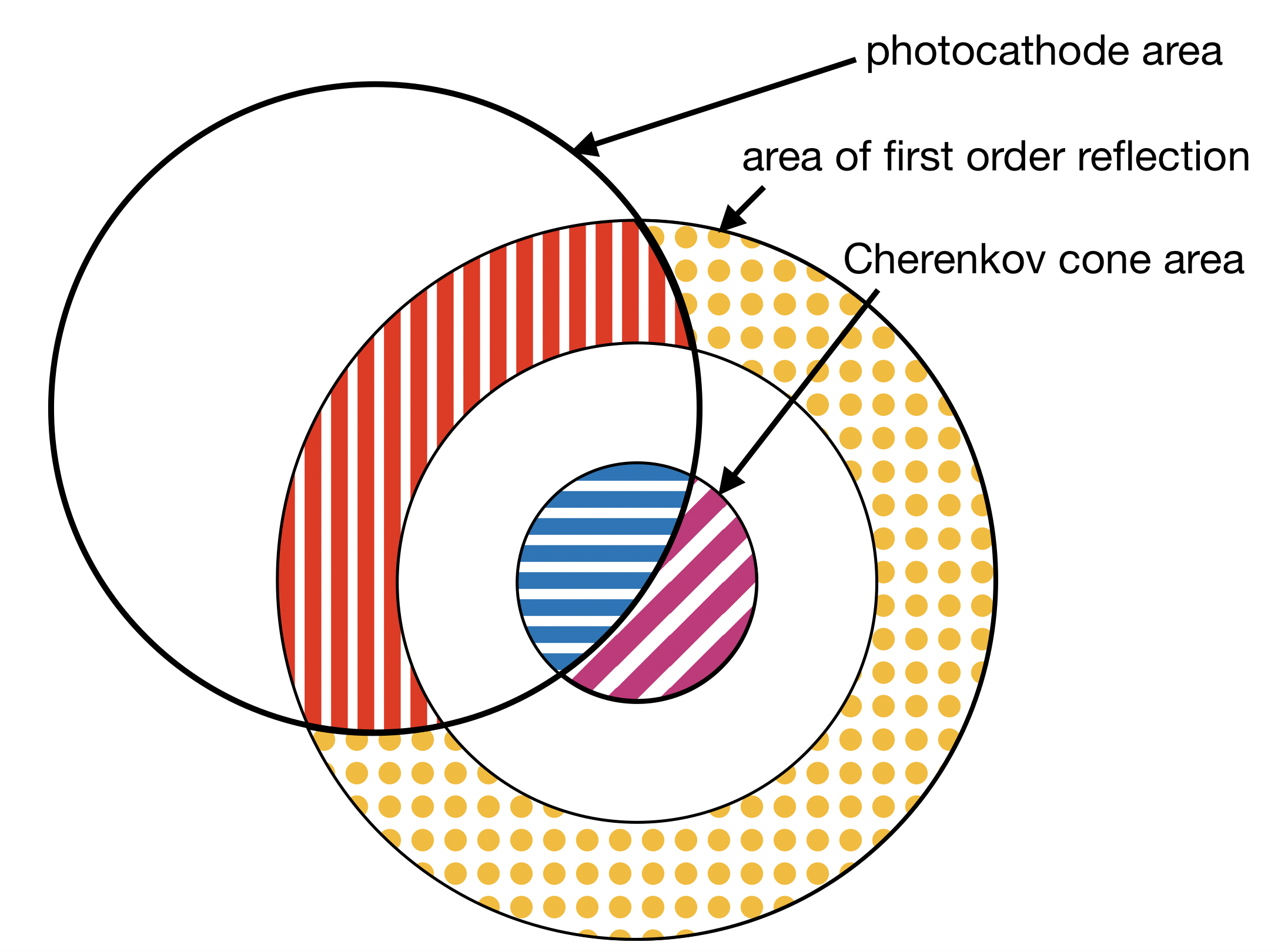}
	  	   		\caption[Geometrical calculation]{Sketch of the geometrical overlap between Cherenkov cone, in blue horizontal hatching and violet diagonal hatching, and first-order reflected light, in red vertical hatching and yellow dotted shaded areas, with the photocathode area, as a thick black line. 
	  	 
	  	   		} 
	  	   		\label{fig:geometrical}
	  	   	\end{figure}

Figure~\ref{fig:geometrical} shows a sketch of the geometrical overlap of the photocathode with the Cherenkov light. In this example, the perpendicularly incident particle is hitting the edge of the photocathode area (thick black line). The blue area (horizontal hatch) shows the Cherenkov light directly reaching the photocathode and the red area (vertical hatch) shows the light from the first-order reflection reaching the photocathode. This light will have survived one reflection on the photocathode and one on the opposite side of the radiator crystal before reaching the photocathode. A part of the light can be transmitted and lost at each reflection. This loss is modelled by a weighting factor ($w<1$) when estimating the amount of light reaching the red area (vertical hatch). 

The amount of light reaching the photocathode decreases with increasing particle impact radius with respect to the photocathode center, due to the decreasing geometrical overlap. With this geometric model the relative amount of photons reaching the photocathode depending on the radius ($P_\text{rel}(r)$) is calculated by:

\begin{equation}
P_\text{rel}(r)=\frac{A_\text{dir.}(r)+w\cdot A_\text{ref.}(r)}{A_\text{dir.}(0)+w\cdot A_\text{ref.}(0)}
\label{relativegeometrical}
\end{equation}
where $A_\text{dir.}(r)$, the area of direct light, and $A_\text{ref.}(r)$, the area of reflected light, depending on the radius, and $w$ is the weighting factor for the loss of the reflected light. This function is scaled to the mean signal charge in the center of the MCP-PMT and $w$ is a free parameter in the fit to the data.

\subsection{Monte-Carlo Simulation of the Light Propagation}
The former analytical geometrical calculation describes the radial behavior of the signal amplitude well for small and large radii but is lacking precision for medium radii. Therefore, a simulation of the light propagation and conversion to charge in the fused silica radiator has been carried out to model the radial amplitude behavior. For this simulation the photons are created as two dimensional points on the window surfaces as objects in a C++ program. A random generator with a probability threshold decides, for each point, if the photon is reflected on the surface or not. If it is reflected, the new position on the opposite side of the window with respect to the Cherenkov angle is calculated. In the initial step, $25\cdot10^6$ points have been randomly distributed over the radiator surface that carries the photocathode. Photons in the photocathode region can be either reflected or, if not reflected, they can be converted to charge or absorbed and thus lost. Outside the photocathode region, photons are lost after not being reflected. The same holds for photons on the opposite side of the radiator, i.e.~at the air-radiator interface. All these effects have been taken into account for each point in the simulation. 

The results of this simulation give the x and y coordinates of the point where each photon is generated as well as where it leaves the crystal, either by transmission or absorption on the crystal surface. Another indicator is given for each photon if it has generated a charge in the photocathode, which means that the photon has not been reflected or absorbed, and ends on the area of the photocathode.

The simulation is controlled by the three probabilities for reflection at the photocathode, absorption at the photocathode if not reflected, and reflection at the air-radiator interface. These parameters have been determined from a $\chi^2$ minimization between simulation results and data.

\subsection{Modeling of the Radial Dependence of Time Resolution} 

The simulation of light propagation has been used as one possible model to quantitatively describe the radial dependence of the time
resolution $\sigma_\text{t}$ on the produced charge. 
It is based on the relation:
\begin{equation}
\sigma_\text{t}^2\propto\frac{\sigma_\text{SPTR}^2}{N_\text{P.E.}}
\label{inami}
\end{equation}

where $\sigma_\text{SPTR}$ is the single photoelectron time resolution of the detector and $N_\text{P.E.}$ is the number of photoelectrons generated at the photocathode. This relation has been shown to be valid for MCP-PMTs, albeit without considering the spatial dependence \citep{Inami:2006cp}. This relation will be extended by including, spatially resolved, absorption and reflection of the Cherenkov photons at the air-radiator interface and at the photocathode.

As the exact quantum efficiency of the photocathode is unknown, we do not extract the number of produced photoelectrons from the simulation. Instead, the ratio between the generated photons and photons reaching the photocathode ($N$) and the ratio of generated photons and those reaching the photocathode after x reflections ($N_\text{x}$) has been used. The number of photoelectrons is linearly correlated to the amount of light reaching the photocathode. Therefore, it is valid to substitute this parameter and rescale Eq.~\eqref{inami} to show the correlation between the time resolution and the relative amount of light reaching the photocathode. 

A distinction between photons reaching the photocathode directly ($N_\text{0}$) and those reaching the photocathode after exactly one
reflection ($N_\text{1}$) is made. The variance of the timing of one MCP-PMT is calculated by

 \begin{equation}
\sigma_\text{MCP1}^2=\left(\frac{N_\text{0}}{N}\right)^2\frac{A}{N_\text{0}}+\left(\frac{N_\text{1}}{N}\right)^2\frac{B}{N_\text{1}}
\label{var1}
\end{equation}

where $N$ is the ratio of all simulated photons to the ones reaching the photocathode, $N_\text{0}$ and $N_\text{1}$ are the ratio of photons reaching the photocathode after 0 or 1 reflections, respectively; $A$ is a scaling factor correlated to $\sigma_\text{SPTR}^2$, $A=(\sigma_\text{MCP1}^2/N_\text{0})_{(r=0)}$ in the center of the MCP-PMT, and $B$ is the corresponding scaling factor for the reflected photons. The reflected photons will increase the signal arrival time jitter and may thus slightly worsen the timing resolution. $B$ is defined as:

 \begin{equation}
B=\left(\sqrt{A}+\Delta\sigma\right)^2
\label{b}
\end{equation}
  
with the additional parameter $\Delta\sigma$, introduced to this model to describe the impact of the reflected photons on the rising edge of the signal and therefore the signal arrival time (SAT). 
	  	   	
\section{Description of the Beam Measurement Setup}

The MCP-PMTs were operated at a nominal gain of $8\cdot10^4$ along with the PICOSEC-Micromegas detectors in muon beams at the CERN SPS secondary beam lines \cite{1994NIMPA.343..351D}. The energy of these muons during our measurements is up to 180\,GeV and the particle rate reached up to $2\cdot 10^7\,s^{-1}$.
The MCP-PMTs feature an unsegmented anode and thus cannot provide position information for the incident particle. Consequently, a beam telescope with three triple-GEM detectors \cite{Sauli:1997qp} with a two-dimensional strip readout structure has been used to track the incoming muons and determine the impact point of the particles on the MCP-PMTs. Details of track reconstruction can be found in \cite{Bortfeldt:2014vvt}. 
The Scaleable Readout System (SRS) \cite{Martoiu:2013sm}, interfacing APV25 front-end boards \cite{FRENCH2001359}, has been used to record GEM detector hits, while the MCP-PMT signals were acquired with oscilloscopes with integrated sampling digitizer with 2.5\,GHz bandwidth at 20\,GS/s sampling rate ("Teledyne LeCroy - WaveRunner 8254"). A common trigger signal, generated by coincident hits in three scintillation detectors, was fed to the SRS and the sampling oscilloscopes. A reliable synchronization of the individual streams was possible, by recording an internally generated SRS trigger number as analog bit stream on an additional oscilloscope channel.

\subsection{Signal Reconstruction and DAQ Calibration}
The time resolution of a detector is defined by the uncertainty of its signal arrival time (SAT) to a time reference. The SAT information is derived by an offline 40\,\% constant fraction discrimination (CFD) analysis for each acquired waveform. The rising edge of each signal has been fitted with a generalized logistic function and the time value of 40\,\% of the amplitude has been calculated for the CFD analysis. Signals of two MCP-PMTs are recorded on two separate channels and their time resolution is evaluated from the difference of their SATs.
Even though both MCP-PMTs are operated at the same amplification voltage, their response is slightly different. For charged particles hitting the innermost region (radius r$<$2.5\,mm), the mean signal charge is $(1.66\pm0.03)$\,pC and $(1.17\pm 0.03)$\,pC, for MCP-PMT 1 and MCP-PMT 2, respectively. Because of different responses, the MCP-PMTs can have different time resolution. Therefore, throughout this study, only the combined time resolution of the two MCP-PMTs is estimated, see Sec.\,\ref{sec:timeresol}.

The contribution of readout electronics and CFD algorithm to the observed time resolution is estimated by recording an identical signal on two oscilloscope channels and determining the variance in the signal arrival time. An 18\,GHz power divider was used to split the signal of one MCP-PMT into two identical signals and no impact on the signal quality could be observed. The time difference of the SAT of both signals follows a distribution with a width of $\sigma$=(3.09$\pm$0.04)\,ps. This leads to a combined timing jitter of the read-out electronics and the CFD algorithm of $\sigma_\text{DAQ}=(3.09\pm0.04)\,\text{ps}$, which yields a time resolution per channel of $\sigma_\text{DAQ}/\sqrt{2}=(2.18\pm0.03)\,\text{ps}$. The measured instrument response function agrees with the trigger and interpolation jitter of $\leq 2.5\,\text{ps}$ RMS expected for the oscilloscope \citep{lecroy}.

\section{Results}

\subsection{Spatial Dependence of the Signal Amplitude} 
\begin{figure}[!htbp]   
	    \subfloat[MCP-PMT 1]{%
    \includegraphics[width=\linewidth]{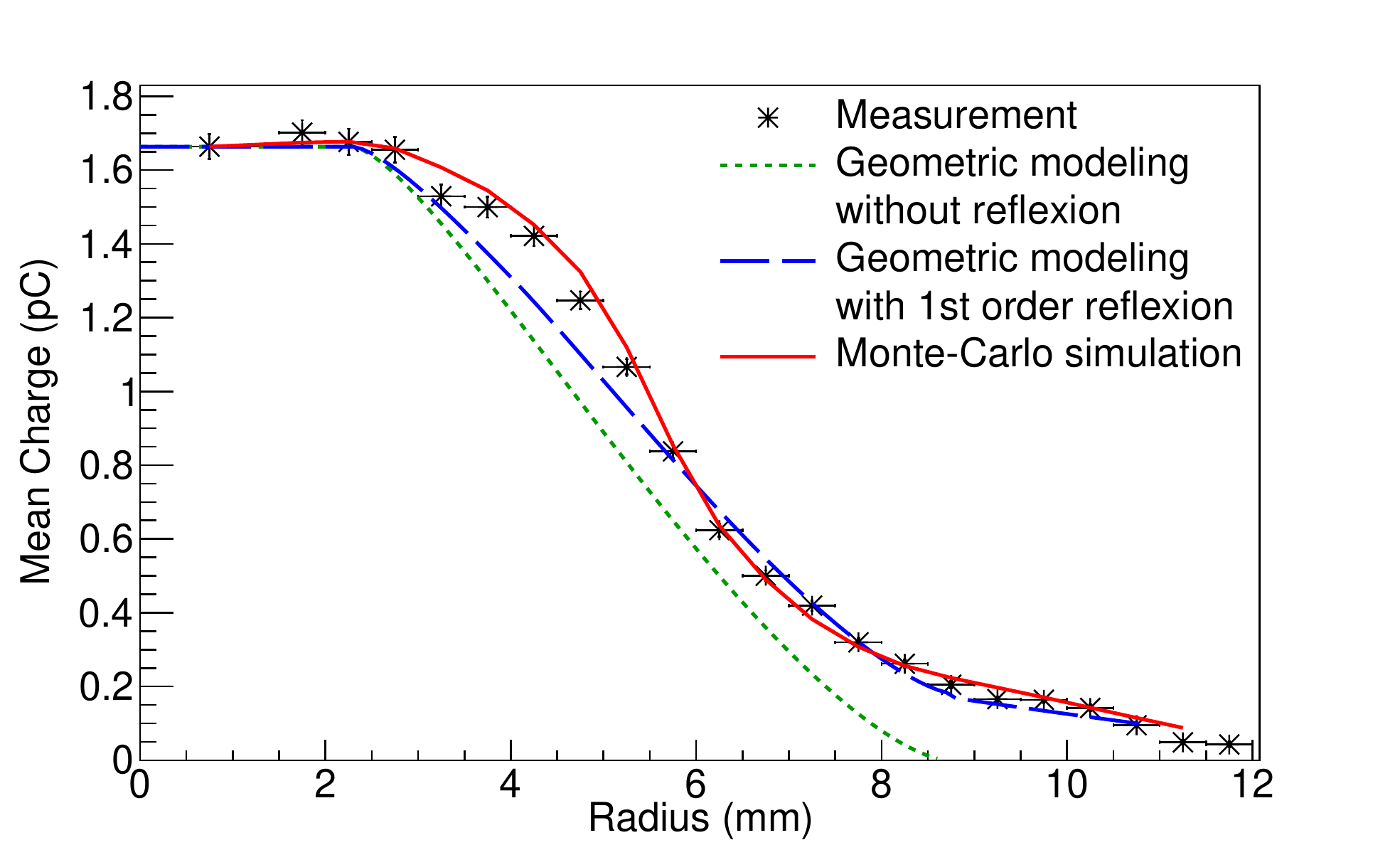}

        }
        \newline
        \subfloat[MCP-PMT 2]{%
           \includegraphics[width=\linewidth]{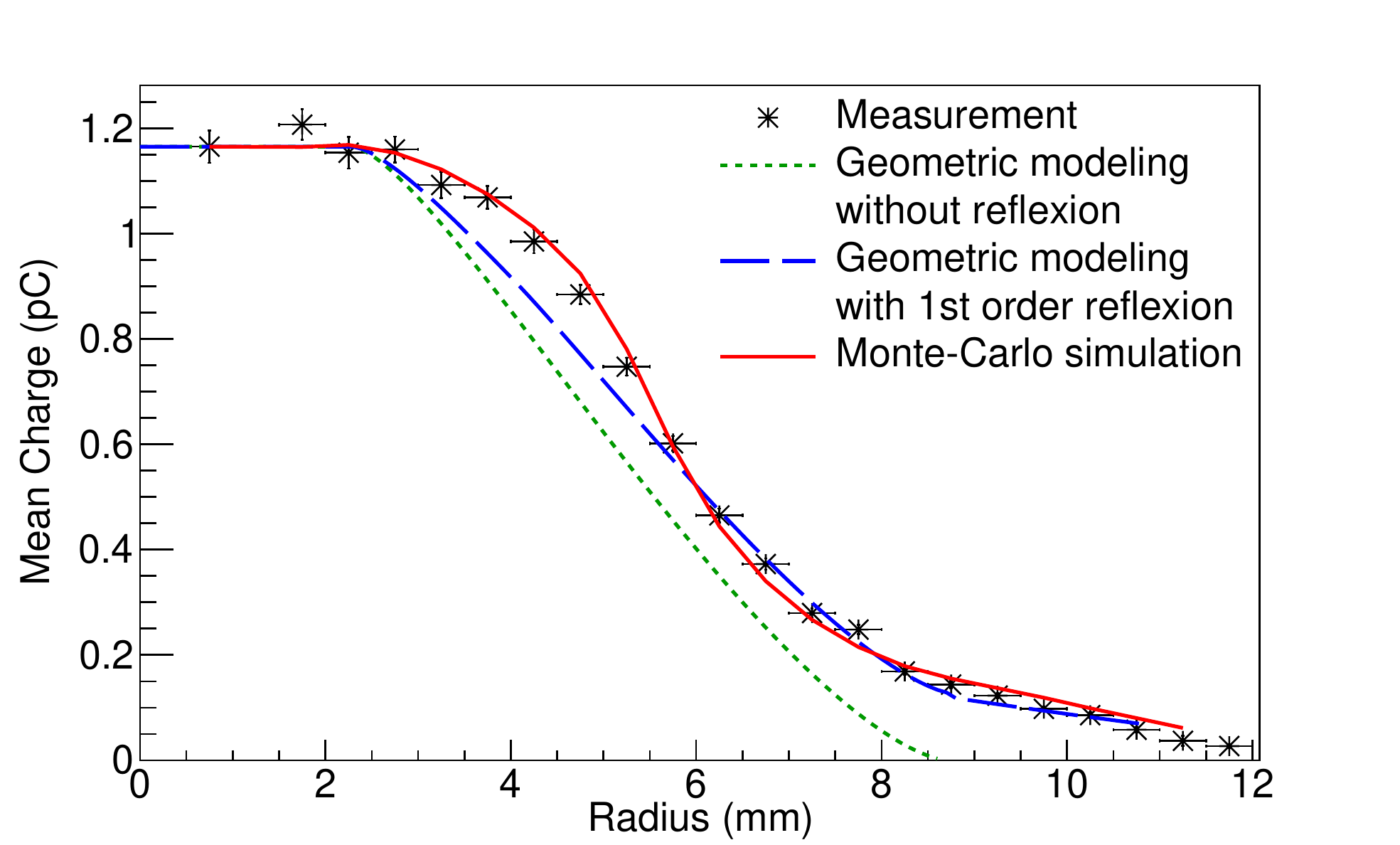}

        }
  		\caption[Charge density Distribution]{Radial distribution of the average signal charge for the two MCP-PMTs. Both MCP-PMTs are operated at 2800\,V. The green dotted curve shows the expected distribution by geometrical calculation without including reflections. The blue dashed line shows the geometrical calculation after including first order reflections. The red line shows the distribution expected from a dedicated simulation that includes the detector characteristics (see text).

        }
	  	\label{fig:chargedis}
	 \end{figure}

Combining signal charge and track impact point measurements, the spatial dependence of the mean signal charge has been evaluated. The radial dependence of the mean signal charges relative to the center of each MCP-PMT are shown in Figure~\ref{fig:chargedis} together with the results from the analytical model with and without reflections and from the Monte-Carlo simulation. The signal charge decreases as expected with increasing distance from the MCP-PMT center due to a decrease of the Cherenkov light reaching the photocathode. Maximum average signal charge is observed for tracks with the full cone contained in the acceptance of the photocathode. This can be observed in the inner $2.5$\,mm of the distribution, as expected for a Cherenkov cone radius of approximately 3\,mm and a photocathode radius of 5.5\,mm. Above $r>2.5$\,mm the mean signal charge starts decreasing, as part of the light coming directly from the Cherenkov cone starts falling outside the photocathode (see Figure\,\ref{fig:window} in Sec.\,\ref{sec:sim}). At radii above 8.5\,mm, only reflected light reaches the photocathode.

The green dotted line in Figure\,\ref{fig:chargedis} shows the overlap between the Cherenkov cone and photocathode without considering any reflection, using the geometric modelling and scaled to the mean signal charge at $r=0$; 
it shows an agreement with the data for small radii, but it underestimates the signal charge for events at larger radii. For the blue dashed line, first-order reflections are added to the geometrical calculation, with fitting the weighting factor to $w=0.08$; the curve shows an agreement with the data in the outer part of the window ($r>$6\,mm). In this region the reflected light dominates the signal.

The red curve shows the result of the Monte-Carlo simulation. From a $\chi^2$ minimization the reflection probability on the photocathode is determined to be $(0.2\pm0.03)$, the absorption probability at the photocathode is $(0.4\pm0.05)$ and the reflection probability at the air-radiator interface is $(0.8\pm0.002)$. All of these parameters are correlated with each other.

For both MCP-PMTs, the radial distribution can be described with the same model parameters by scaling the model output to the different mean signal charges in the center of the photocathode. The determined parameters need to be taken with a grain of salt. More refined models with precise knowledge on the MCP-PMT materials and including signal formation processes might yield different results.

\subsection{Time Resolution}\label{sec:timeresol}

\begin{figure}[!htb]
  \centering
  \includegraphics[width =  \linewidth]{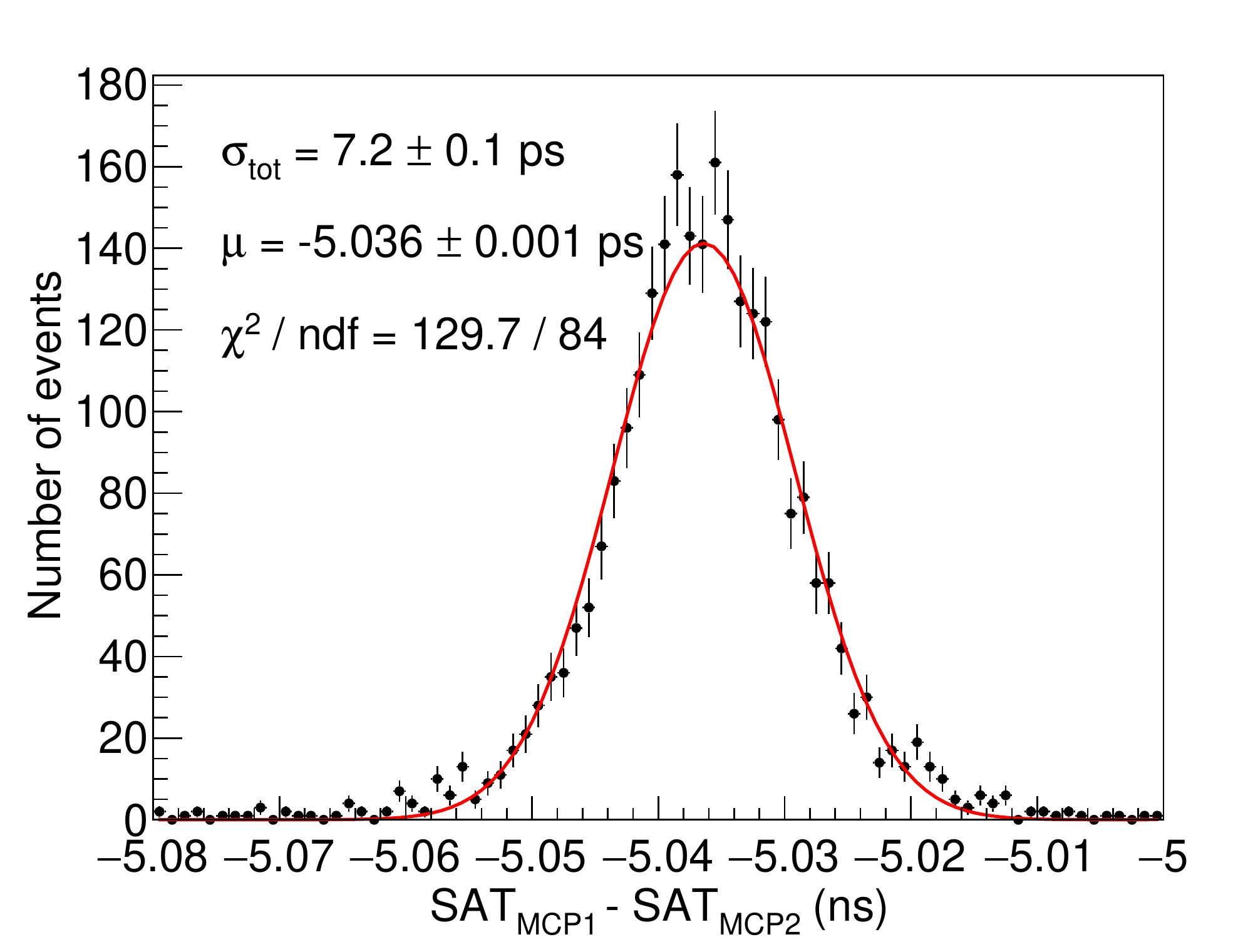}
    \caption[]{Difference of the signal arrival times at the two MCP-PMTs. Data are selected by requiring a particle track passing through the inner 11\,mm of the first MCP. A combined time resolution of (7.2$\pm$0.1)\,ps is measured from a Gaussian fit (red line).}
  \label{singlemcp}
\end{figure}

The time difference of the SAT of both MCP-PMTs results in a Gaussian-like distribution (Figure~\ref{singlemcp}). 
The standard deviation $\sigma_\text{tot}$ of this distribution is determined from fitting a Gaussian function and
\begin{equation}
\sigma_\text{tot}= \sqrt{\sigma_\text{MCP1}^2 +\sigma_\text{MCP2}^2 +\sigma_\text{DAQ}^2}~,
\label{eq:sigmatot}
\end{equation}
where $\sigma_\text{MCP1}$ and $\sigma_\text{MCP2}$ are the time resolutions of MCP-PMT 1 and MCP-PMT 2, respectively, and $\sigma_\text{DAQ}$ denotes the setup-specific contribution of DAQ system and analysis method.

An extraction of the time resolution of a single MCP-PMT from these beam measurements is only possible when assuming that $\sigma_\text{MCP1} \equiv \sigma_\text{MCP2}$ at the same track proximity to the center. As the respective alignment of the two MCP-PMTs is measured to be $(0.38\pm0.01)$\,mm, as the 2D-distance between the MCP-PMT centers, and they are operated at equal high voltage, this assumption is plausible. However, the observed pulse height in MCP-PMT 2 is 30\,\% smaller than that in MCP-PMT 1. This could be due to a different photo-conversion efficiency or amplification factor. It has to be assumed that the dependence of the resolution on the signal amplitude is the same despite the observed gain difference, to extract the time resolution of a single MCP-PMT. Since the available dataset does not allow for a determination of these effects on the time resolution, we will only discuss in the following the combined MCP-PMT time resolution:

\begin{equation}
    \sigma_\text{MCP} \coloneqq \sqrt{\sigma_\text{MCP1}^2 +\sigma_\text{MCP2}^2} = \sqrt{\sigma_\text{tot}^2-\sigma_\text{DAQ}^2}~.
    \label{eq:sigmasingle}
\end{equation}

If the two MCP-PMTs behaved similarly, a time resolution of $\sigma_\text{MCP}/\sqrt{2}$ would be expected for each MCP-PMT individually.

\subsection{Spatial Dependence of Time Resolution} 
\begin{figure}[!htb]
	  	   		\centering
	  	   		\includegraphics[width =  \linewidth]{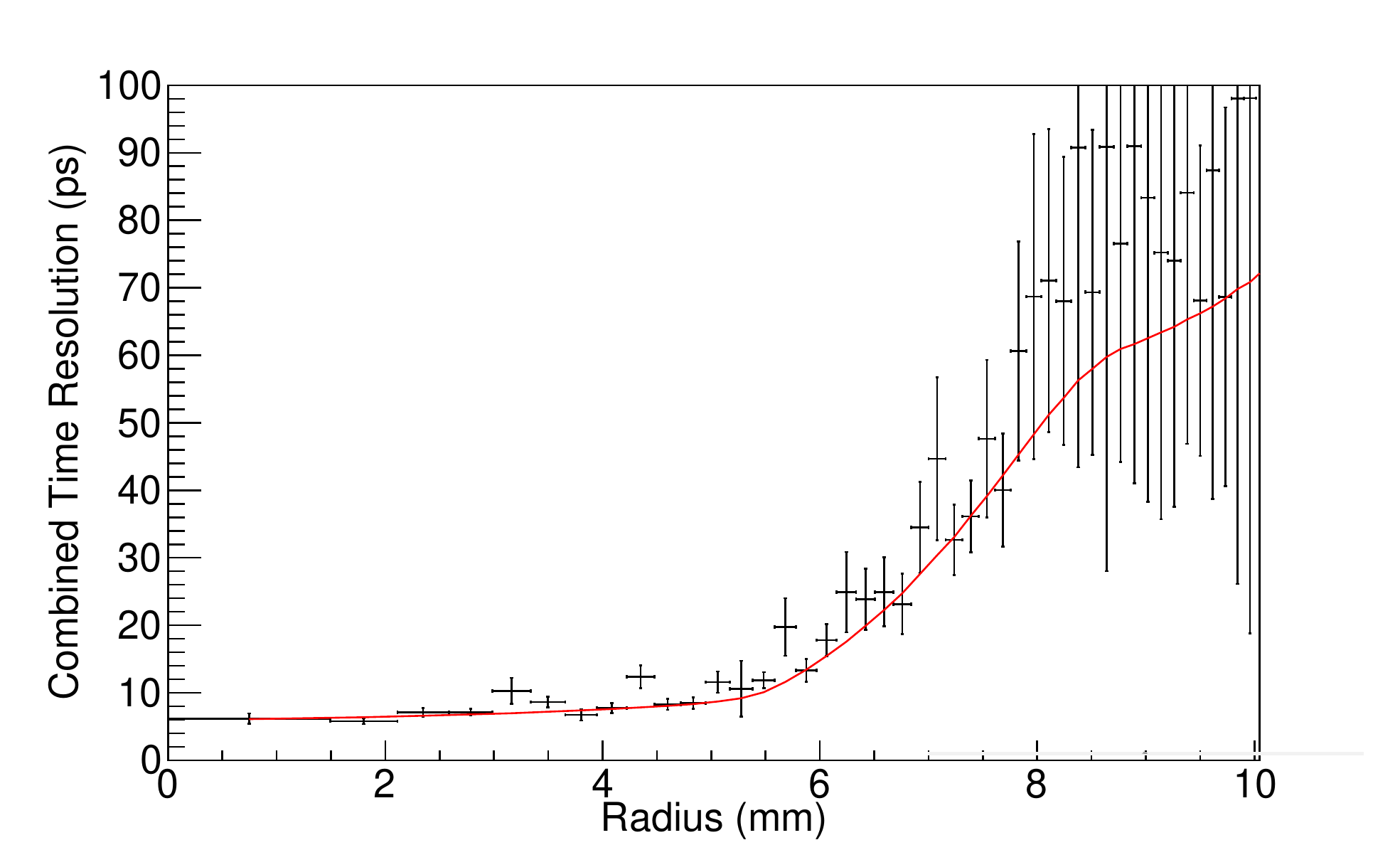}
	  	   		\caption[Spatial distribution of the time resolution]{Combined time resolution as a function of the track impact point distance from the photocathode center. The red line shows a possible modelling using results from the Monte-Carlo simulation, as explained in Sec.\,\ref{sec:sim}.
	  	   		}
	  	   		\label{fig:timedis}
	  	   	\end{figure}

In the following, we will discuss the time resolution of the MCP-PMTs as a function of the impact point of the incident particle. Events are grouped according to their impact radius such that the statistics of events hitting each ring-like area is equal and the time resolution is then independently calculated for each group. Figure~\ref{fig:timedis} shows the combined time resolution $\sigma_\text{MCP}$ as a function of the distance with respect to the center of MCP-PMT 2.

The combined time resolution is better than $10$\,ps in the inner radius of r$<$4\,mm. Beyond that, the time resolution starts gradually degrading up to several tens of picoseconds.

This result is consistent with the measured signal mean charge. In the inner radius of 2.5\,mm, the full Cherenkov light cone is entirely projected on the photocathode and the signal pulse amplitude is maximal in this region. As the impinging particles are further away from the center of the MCP-PMT, the mean signal charge is reduced and the time resolution degrades.

The red curve in Figure~\ref{fig:timedis} shows the results of the Monte-Carlo simulation (see Eq.\,\eqref{var1}) described in Sec.\,\ref{sec:sim}. Optimal agreement between simulation results and observed data is reached with A=6.69\,ps$^2$, $\Delta\sigma=7.5\,\text{ps}$, and B=101.75\,ps$^2$.
The additional spread for the reflected photons of both MCP-PMTs $\Delta\sigma$ must not be confused with the time delay of the reflected photons, which is of the order of $\sim$ 40 to 45\,ps. $\Delta\sigma$ depends on the probabilities for reflection and absorption used in the Monte-Carlo model of the charge distribution.

\section{Conclusions}
A study of the MCP-PMT timing resolution was performed using muons with energies up to 180\,GeV in a test beam at CERN. In particular, the study aimed at characterizing the time resolution as a function of the position of the impinging muons. The MCP-PMT provides a precise and reliable reference time well below 10 ps for charged particles, traversing the MCP-PMT inside the photocathode area. However, when the particle traverses the detector further outside the photocathode area, the timing resolution begins to degrade due to a decrease of the detected fraction of Cherenkov light.

The model, based on simple geometrical arguments, has shown good agreement with the observed behavior by assuming an additional time jitter of 7.5\,ps due to the reflected light. We conclude that the MCP-PMT is appropriate for use as a time reference to determine the time resolution of detector prototypes like the PICOSEC-Micromegas, that have shown a time resolution of 25 to 30\,ps \citep{Bortfeldt:2017jb}, as long as the investigated region in the detector under test is well aligned to the MCP-PMT and has a diameter smaller than 13\,mm. 

\section*{Acknowledgments}
We acknowledge the financial support of the Cross-Disciplinary Program on Instrumentation and Detection of CEA,
the French Alternative Energies and Atomic Energy Commission; the RD51 collaboration, in the framework of RD51 common projects; and
the Fundamental Research Funds for the Central Universities of China.
L.~Sohl acknowledges the support of the PHENIICS Doctoral School Program of Universit\'e Paris-Saclay.
J.~Bortfeldt acknowledges the support from the COFUND-FP-CERN-2014 program (grant number 665779).
M.~Gallinaro acknowledges the support from the Funda\c{c}\~ao para a Ci\^{e}ncia e a Tecnologia (FCT), Portugal
(grants IF/00410/2012 and CERN/FIS-PAR/0006/2017).
F.J.~Iguaz acknowledges the support from the Enhanced Eurotalents program (PCOFUND-GA-2013-600382).
S. White acknowledges partial support through the US CMS program under DOE contract No. DE-AC02-07CH11359.

\bibliographystyle{JHEP}
\bibliography{biblio}

\end{document}